\documentclass[pra,twocolumn,showpacs,10pt]{revtex4}
\usepackage{graphicx}% Include figure files
\begin{document}

\title{Entanglement between two fermionic atoms inside a cylindrical harmonic trap}

\author{B. Sun and L. You}
\affiliation{School of Physics, Georgia Institute of Technology,
Atlanta, GA 30332, USA}
\author{D. L. Zhou}
\affiliation{Institute of Physics, Chinese Academy of Sciences, Beijing 10080, China}

\begin{abstract}
We investigate quantum entanglement between two (spin-1/2)
fermions inside a cylindrical harmonic trap, making use of the von
Neumann entropy for the reduced single particle density matrix as
the pure state entanglement measure. We explore the dependence of
pair entanglement on the geometry and strength of the trap and on
the strength of the pairing interaction over the complete range of
the effective BCS to BEC crossover. Our result elucidates an
interesting connection between our model system of two fermions
and that of two interacting bosons.
\end{abstract}


\maketitle

\section{introduction}
Recent experiments with lattice fermions across Feshbach resonance
have raised significant hope for the application of atomic quantum
gases to implementations of quantum information processing
\cite{hulet,ETH,ETH2}. These experiments usually begin with the
preparation of two fermions into each optical lattice site. Making
use of a Feshbach resonance \cite{fr}, the two-atom scattering
length is tuned from a small positive value to the attractive side
by varying an external magnetic field. During this process, the
motional states of the fermionic atom pair in each lattice site
can occupy different Bloch bands, exhibiting new features that
need further theoretical explanation \cite{JasonHo}.

In quantum information science, quantum entanglement is viewed as
the enabling resource for its potential power over classical
information processing, as in quantum teleportation
\cite{tele1,tele2,tele3,tele4,tele5} and cryptography
\cite{schumacher1,qc1,qc2}. In this paper, we present a thorough
investigation of the pair entanglement of two fermions in a single
optical lattice site, approximated as a cylindrical harmonic trap.
Similar studies have been carried out in a one dimensional
harmonic trap \cite{sun} and in a three dimensional spherical
harmonic trap \cite{law}. The present study differs in two
significant aspects: first, we extend earlier studies to the case
of three dimensional cylindrical harmonic traps; second, we allow
for the possibility that the two interacting atoms may form a
molecular bound state, albeit in the broad resonance regime. In
the model to be presented below we will parameterize the two atom
interaction using a general formalism commonly adopted in the many
body system of BCS to BEC crossover
\cite{jin,grimm,ketterle,hulet2}. For a broad Feshbach resonance,
where most of the currently employed resonance theories do fall
into, the earlier results based on a single channel model are
easily recovered by excluding the molecular component.

This paper is organized as follows. First, we present the model
system and our formulation, essentially parallel to the
development and notations of Ref. \cite{JasonHo}. Next, we focus
on the broad resonance regime and briefly discuss the molecular
component. This paves the way for the discussion of our central
result --- a thorough investigation of the pair entanglement and
its dependence on trap strength and geometry, and on atom-atom
interaction strength. Finally, we conclude with a brief summary.

\section{Two interacting fermionic atoms in a harmonic trap}
Following the successful model proposed in Ref. \cite{JasonHo},
two fermions are assumed to be located in the same lattice site,
which for this study is approximated as a harmonic trap. Each
optical lattice site represents an independent system, since the
quantum tunnelling, or hopping between neighboring sites, is
negligible for deep optical lattice potential of interest. The
Hamiltonian for the system of two spin-$1/2$ fermionic atoms is
given by
\begin{eqnarray}
H &=& \sum_{\mathbf{m}\sigma} E_{\mathbf{m}}
a^\dagger_{\mathbf{m}\sigma} a_{\mathbf{m}\sigma}  +  \bar{\nu}
b^\dagger b \nonumber\\
&&+ \sum_{\mathbf{m},\mathbf{n}}
\alpha_{\mathbf{m},\mathbf{n}} \left( a^\dagger_{\mathbf{m}\uparrow}
a^\dagger_{\mathbf{n}\downarrow} b + h.c. \right),
\label{hm}
\end{eqnarray}
where $E_{\mathbf{m}} = \sum_{j=x,y,z} \hbar \omega_j(m_j+1/2)$ is
the harmonic oscillator energy for state labelled by
$\mathbf{m}=(m_x,m_y,m_z)$ with angular frequencies
$(\omega_x,\omega_y,\omega_z)$. $a^\dagger_{\mathbf{m}\sigma}$ is
the creation operator for a fermionic atom in the open channel
with energy $E_{\mathbf{m}}$ and spin $\sigma$. $b^\dagger$ is the
creation operator for the two-atom bound state, a bosonic molecule
in the closed channel with its center of mass wave function fixed
exactly at the harmonic ground state, a result of the simple
approximation that the optical lattice trap potential for the
bound state molecule is simply equal to the sum of the trap
potentials of the two atoms. $\bar{\nu}$ is the energy difference
between the closed channel bosonic molecule and the two fermions
in the open channel. $\alpha_{\mathbf{m},\mathbf{n}}$ is the
coherent coupling element converting two open channel fermions
into a closed channel bosonic molecule, which is defined to
contain a common constant prefactor $\alpha$, i.e.,
$\alpha_{\mathbf{m},\mathbf{n}}=\alpha\langle \mathbf{0}_{\rm
c.m.},\psi_{\rm rel}|\mathbf{m},\mathbf{n}\rangle$ with
$|\mathbf{m}\rangle$ being the harmonic orbital of a single atom.
The relative motional part of the molecular ground state,
$|\psi_{\rm rel}\rangle$, will be approximated as a contact
$\delta(\vec r)$ function, since it is typically of atomic size
\cite{JasonHo,heinzen}, much less than other length scales in this
problem. Such a simplification contains an ultra-violet divergence
that can be removed by a suitable momentum cutoff with a
renormalized detuning \cite{heinzen}. An alternative formulation
involves the use of the regularized delta function \cite{moore}.
The cylindrical harmonic trap is characterized by the trap
frequencies $\omega_x=\omega_y=\omega_{\perp}=\omega_z/\lambda$
with $\lambda$ parameterizing the trap aspect ratio. Within this
model, the two fermionic atoms in the open channel (of being
atoms) at bands $\mathbf{m}$ and $\mathbf{n}$ can be converted
into a closed channel bosonic molecule, or vice versa. The
eigenstate of Hamiltonian (\ref{hm}) can be expressed as
\begin{eqnarray}
|\Psi\rangle = \left( \beta b^\dagger + \sum_{\mathbf{m},\mathbf{n}}
\eta_{\mathbf{m},\mathbf{n}} a^\dagger_{\mathbf{m}\uparrow}
a^\dagger_{\mathbf{n}\downarrow} \right)|{\rm vac}\rangle,
\label{2w}
\end{eqnarray}
again following Ref. \cite{JasonHo}. The various coefficients are
determined from the following coupled equations
\begin{eqnarray}
\eta_{\mathbf{m},\mathbf{n}} &=& \beta {\alpha_{\mathbf{m},\mathbf{n}} \over E-E_{\mathbf{m},\mathbf{n}} },\\
E-\bar{\nu} &=& \sum_{\mathbf{m},\mathbf{n}} {\alpha_{\mathbf{m},\mathbf{n}}^2 \over E-E_{\mathbf{m},\mathbf{n}} },
\label{EEq}\\
{1\over \beta^2} &=& 1 + \sum_{\mathbf{m},\mathbf{n}}
{\alpha_{\mathbf{m},\mathbf{n}}^2 \over
(E-E_{\mathbf{m},\mathbf{n}})^2 }.
\end{eqnarray}
As in Ref. \cite{JasonHo}, the divergence in Eq. (\ref{EEq}) can
be removed by introducing a cutoff in the summation. Such a
procedure will renormalize $\bar{\nu}$ to $\nu^*$. By comparing
the free space expression of the energy $E$ (renormalized) to the
results of resonance scattering at low energies, the parameters
$\nu^*$ and $\alpha$ are then matched to the experimentally
relevant parameters $a_s$ and $r_0$, where $a_s$ is the $s$-wave
scattering length between fermionic atoms in different internal
states and $r_0$ is the effective range. The result is
\cite{JasonHo}
\begin{eqnarray}
{1\over a_s}=-{\sqrt{2}\nu^*\hbar^2\over m\alpha^2
\pi^{7/2}}=-{\nu^*|r_0|m\over 2\hbar^2},
\end{eqnarray}
where $m$ is the mass of the atom. A dimensionless parameter
$x=E/(2\hbar\omega_{\perp})-1-\lambda/2$ is then introduced, with
which the energy quantization condition becomes
\begin{eqnarray}
\sqrt{2\lambda} \left[{d_{\perp}\over a_s} + {|r_0|\over
d_{\perp}} \left(x+1+{\lambda\over 2}\right)\right] = -
{\lambda\over \sqrt{\pi}} F\left(-{x\over \lambda},{1\over
\lambda}\right)\label{qzc},
\end{eqnarray}
where $d_{\perp}\equiv\sqrt{\hbar/(m\omega_\perp)}$, and the
function on the right hand side of Eq. (\ref{qzc}) is defined in
Ref. \cite{TCalarco}
\begin{eqnarray}
F(u,\eta)=\int_0^\infty dt \left( {\eta e^{-ut}\over
\sqrt{1-e^{-t}}(1-e^{-\eta t}) } -{1\over t^{3/2}}\right).
\end{eqnarray}
For a spherical trap this reduces to the well-known result of
$F(-x,1)=-2\sqrt{\pi}\,\Gamma(-x)/\Gamma(-x-1/2)$ \cite{TBush}.
Furthermore, if $r_0= 0$, the energy spectrum and eigenfunctions
coincide with the results of two atoms in a harmonic trap as
studied previously in Refs. \cite{TBush,TCalarco}. This is a
straightforward conclusion, since for $r_0=0$, the current model
describes the same physical process quantified by a single
$s$-wave parameter $a_s$ \cite{TBush,TCalarco}. A non-zero $r_0$
as incorporated in Ref. \cite{JasonHo} allows for a more general
model including both open channel fermions and a closed channel
bosonic molecule.

\begin{figure}[h]
\centering
\includegraphics[width=3.25in]{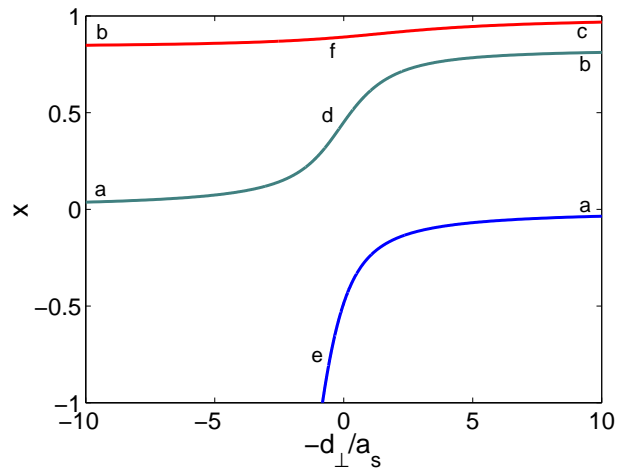}
\caption{(Color online) Energy spectrum of two fermions in a
cylindrical harmonic trap versus inverse scattering length in the
broad resonance regime. Here we choose $\lambda=5/6$ and
$|r_0|/d_{\perp}=0.04$. } \label{energy}
\end{figure}

The energy spectrum versus inverse scattering length
$-d_\perp/a_{s}$ for the current model is shown in Fig.
\ref{energy}. Across a Feshbach resonance, the scattering length
can be tuned according to $a_s=a_{\rm bg}(1-\Delta/(B-B_0))$
\cite{frt}, where $a_{bg}$ is the background scattering length.
$B_0$ and $\Delta$ are the resonant field and width, respectively.
The effective range $r_0$ is found to satisfy $r_0=-2\hbar^2 /(
m\mu a_{bg}\Delta)$ \cite{eff}. $\mu$ is the magnetic moment
difference in the open and closed channel.

%obtained from the definition $k\cot(\delta_0(k)) = -1/a_s + r_0
%k^2/2$ for $s$-wave collision phase shift $\delta_0(k)$ at energy
%$\hbar^2k^2/(2m)$ \cite{eff}.

In this work when dealing with a broad Feshbach resonance, the
validity of our model calculation requires that $|r_0|\ll
d_{\perp}, |a_s|$. As in previous studies \cite{TBush,TCalarco},
the eigenenergy remains an increasing function of the inverse
scattering length $-d_\perp/a_{s}$. In a more complete treatment,
the closed molecular channel is observed to differ from the lowest
two atom (bound) state $|$e-a$\rangle$ as shown in this figure in
the internal spin state \cite{frt}.

\section{Results on pair entanglement}
In this study, we will limit our discussion to the pair
entanglement for a broad Feshbach resonance, corresponding to the
regime of $|r_0|\ll d_\perp$. The nonclassical correlations
(squeezing and entanglement) were previously considered for
non-condensate atoms across a Feshbach resonance in free space
\cite{ben}. In the following we shall therefore focus on the two
adiabatic eigenstates labelled as $|$a-d-b$\rangle$ and
$|$b-f-c$\rangle$ in Fig. \ref{energy}.

First, we briefly review the result on the molecular component
according to Ref. \cite{JasonHo}. Starting from a small positive
scattering length, when adiabatically following the state
$|$a-d-b$\rangle$, the molecular component $P_{\rm
mol}\equiv\beta^2 \ll 1$ remains small in the broad resonance
regime \cite{ho2,JasonHo}. More precisely,
\begin{eqnarray}
\beta^2 =
{|r_0|\over d_{\perp}} {\partial x \over
\partial \left(-{d_{\perp}\over a_s}\right) },
\end{eqnarray}
i.e., the molecular component is always small because of the
small pre-factor ${|r_0|/ d_{\perp}}$. Even smaller $P_{\rm  mol}$
is expected along the state $|$b-f-c$\rangle$ because it has a
weaker dependence on the x-axis as shown in Fig. \ref{energy}.
Numerically we find that the molecular population remains less
than $1\%$ for $|r_0|/d_{\perp}=0.04$.

In a spherical harmonic trap, the entanglement properties for
$r_0=0$ have already been studied before \cite{law}. Making use of
our model as outlined above based on Ref. \cite{JasonHo}, we now
extend the earlier result \cite{law} to a cylindrical harmonic
trap. To begin with, we first approximate the two fermion wave
function (\ref{2w}) by neglecting the small molecular component,
thus we obtain
\begin{eqnarray}
|\Psi\rangle = \sum_{\mathbf{m},\mathbf{n}}
\eta_{\mathbf{m},\mathbf{n}} a^\dagger_{\mathbf{m}\uparrow}
a^\dagger_{\mathbf{n}\downarrow} |{\rm vac}\rangle,
\end{eqnarray}
with the normalization constraint
$\sum_{\mathbf{m},\mathbf{n}}|\eta_{\mathbf{m},\mathbf{n}}|^2 =
1$. $\eta_{\mathbf{m},\mathbf{n}}$ are assumed real and symmetric
for two identical atoms with
$\eta_{\mathbf{m},\mathbf{n}}=\eta_{\mathbf{n},\mathbf{m}}$. In
the single atom basis state, the above wave function becomes
\begin{eqnarray}
|\Psi\rangle = \sum_{\mathbf{m},\mathbf{n}}
\eta_{\mathbf{m},\mathbf{n}}|\mathbf{m}_1 \mathbf{n}_2 \rangle
{|\uparrow_1\downarrow_2\rangle -
|\downarrow_1\uparrow_2\rangle\over \sqrt{2}}.
\end{eqnarray}
Such a state has both spatial and spin degrees of freedom, but the
two degrees of freedom remain factorized. We therefore adopt the
von Neumann entropy as our entanglement measure for a pure state.
The factorized spin degree part simply contributes a
$\mathrm{ln}(2)$, and the total entropy becomes $\mathbf{E} =
\mathbf{E}_{\rm spatial} + \mathrm{ln}(2)$ \cite{neilson}. While
nontrivial, the spatial part of entanglement $\mathbf{E}_{\rm
spatial}$ (abbreviated as pair entanglement) can be computed from
the Schmidt decomposition \cite{schmidt} of the spatial wave
function, i.e., we need to find
\begin{eqnarray}
\sum_{\mathbf{m},\mathbf{n}}
\eta_{\mathbf{m},\mathbf{n}}|\mathbf{m}\rangle_1 |\mathbf{n}
\rangle_2 = \sum_{\mathbf{q}} \kappa_{\mathbf{q}}
|\phi_{\mathbf{q}}\rangle_1 |\psi_{\mathbf{q}}\rangle_2,
\label{sd}
\end{eqnarray}
with the Schmidt mode functions
$|\phi_{\mathbf{q}}\rangle=|\psi_{\mathbf{q}}\rangle$, because
$\eta_{\mathbf{m},\mathbf{n}}$ is symmetric. From this, the von
Neumann entropy is found to be $\mathbf{E}_{\rm spatial} =
-\sum_{\mathbf{q}} \kappa_{\mathbf{q}}^2 \rm{ln}
(\kappa_{\mathbf{q}}^2 )$. Computing the direct Schmidt
decomposition for the three dimensional wave function turns out to
be a quite demanding numerical task. Fortunately we can simplify
this problem effectively to several one dimensional Schmidt
decompositions, which will be discussed elsewhere \cite{bof}.

\begin{figure}[h]
\centering
\includegraphics[width=3.25in]{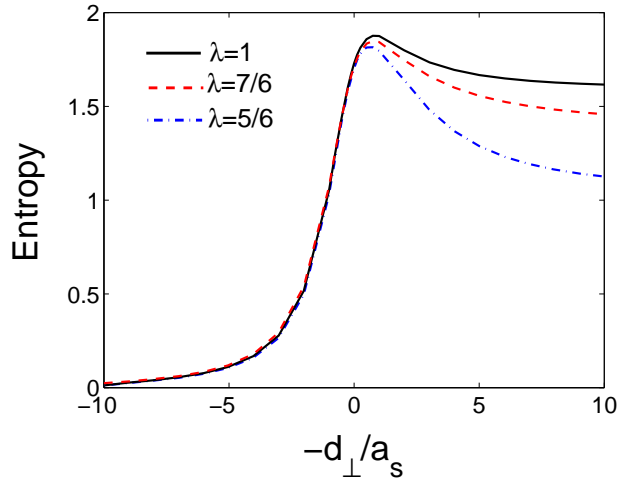}
\caption{(Color online) Pair entanglement versus the atomic
interaction strength at different trap aspect ratio $\lambda$ for
the state $|$a-d-b$\rangle$ at $|r_0| / d_{\perp} =0.04$. The
dependence on $r_0$ is very small within the broad resonance
regime.} \label{ent1}
\end{figure}

\begin{figure}[h]
\centering
\includegraphics[width=3.25in]{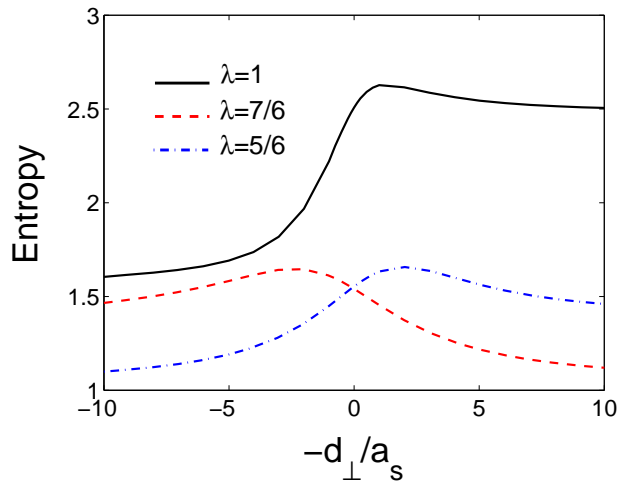}
\caption{(Color online) The same as in Fig. \ref{ent1} except for
state $|$b-f-c$\rangle$.} \label{ent2}
\end{figure}

For the two adiabatic states $|$a-d-b$\rangle$ and
$|$b-f-c$\rangle$, we have numerically evaluated their spatial
pair entanglement. The results are shown in Figs. \ref{ent1} and
\ref{ent2}. For both states, we find that pair entanglement is a
smooth function of $-d_{\perp}/a_s$. The pair entanglement first
increases to some maximal value, then decreases and saturates to
certain finite value. The overall dependence on the atom
interaction strength remains essentially the same as before
\cite{law}.

We now discuss several interesting limits. First, for an
approximately spherical trap with $\lambda\sim1$, the adiabatic
state $|$a-d-b$\rangle$ corresponds to $|a\rangle$ or $|b\rangle$
at $-d_{\perp}/a_s\to -\infty$ or $+\infty$, respectively. For
$\lambda=5/6$, we find $|b\rangle\propto\big[\sum_{n=1,2}
(c_{nz}^\dagger)^2 - 2c_{1z}^\dagger c_{2z}^\dagger \big]
|000\rangle_1|000\rangle_2 $, whose pair entanglement is
$\rm{ln}(2\sqrt{2})\approx 1.04$. Here we again follow the
notation of Ref. \cite{JasonHo} with $c_{nj}^\dag$ the creation
operator for a fermionic atom indexed by $n$ in the $j$-th trap
direction. State $|m_1m_2m_3\rangle_n$ therefore refers to atom
$n$ in the motional state $m_j$ along the $j$-th direction. For
$\lambda=7/6$, we find $|b\rangle \propto\big[ \sum_{n=1,2;j=x,y}
(c_{nj}^\dagger)^2 - 2\sum_{j=x,y} c_{1j}^\dagger c_{2j}^\dagger
\big] |000\rangle_1|000\rangle_2$ with a pair entanglement
$\rm{ln}(4)\approx 1.39$. For $\lambda=1$, we find
$|b\rangle\propto\big[\sum_{n=1,2;j=x,y,z} (c_{nj}^\dagger)^2 -
2\sum_{j=x,y,z} c_{1j}^\dagger c_{2j}^\dagger \big]
|000\rangle_1|000\rangle_2 $ with a pair entanglement
$\rm{ln}(2\sqrt{6})\approx 1.59$. We find that the pair
entanglement is always larger at the limit of a spherical trap
with $\lambda = 1 $. For the adiabatic state $|$b-f-c$\rangle$, we
find generally that the pair entanglement at $\lambda=1$ is well
separated from $\lambda\neq 1$ because of the increased motional
state degeneracy. State $|c\rangle$ corresponds to the limit of
$-d_{\perp}/a_s\to+\infty$, which for $\lambda=7/6$ becomes
$|c\rangle\propto\big[ \sum_{n=1,2} (c_{nz}^\dagger)^2 -
2c_{1z}^\dagger c_{2z}^\dagger \big] |000\rangle_1|000\rangle_2 $,
which is precisely the state $|b\rangle$ for $\lambda=5/6$ in the
adiabatic state $|$a-d-b$\rangle$, whose pair entanglement
therefore remains the same $\rm{ln}(2\sqrt{2}) \approx 1.04$. This
correspondence persists also for $\lambda=5/6$, where the state
$|c\rangle$ is the same as state $|b\rangle$ for $\lambda=7/6$ in
the adiabatic state $|$a-d-b$\rangle$. The state $|c\rangle$ at
$\lambda=1$ is more complicatedly expressed as a linear
combination of different single particle states as
$|c\rangle\propto\big[ -\sum_{n=1,2;j=x,y,z}(c^\dagger_{nj})^4 -
\sum_{n,m;j\neq k}(c^\dagger_{mj})^2 (c^\dagger_{nk})^2-
6\sum_{j}(c^\dagger_{1j})^2 (c^\dagger_{2j})^2 + 4\sum_{n\neq
m;j}c^\dagger_{mj} (c^\dagger_{nj})^3 + 4 \sum_{n;j\neq
k}c^\dagger_{1j} c^\dagger_{2j} (c^\dagger_{nk})^2 - 8 \sum_{j\neq
k}c^\dagger_{1j} c^\dagger_{1k} c^\dagger_{2j} c^\dagger_{2k}
\big]|000\rangle_1|000\rangle_2 $, whose pair entanglement is
$55\rm{ln}(2)/24+7\rm{ln}(3)/8-\rm{ln}(5)/24\approx 2.48$.

\begin{figure}[h]
\centering
\includegraphics[width=3.25in]{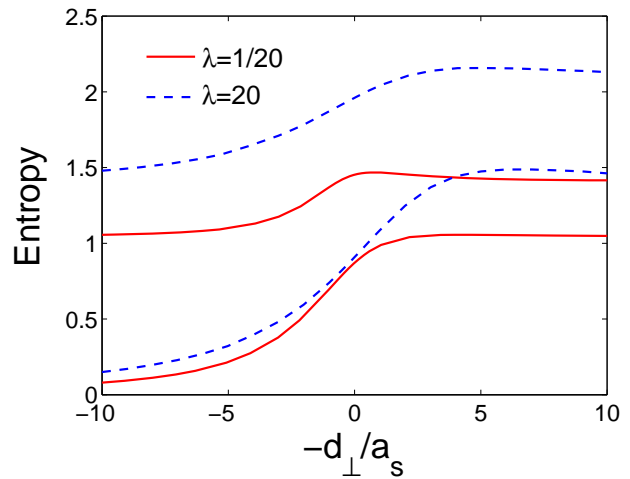}
\caption{(Color online) Entropy as function of inverse scattering
length for processes $|$a-d-b$\rangle$ and $|$b-f-c$\rangle$. Red
solid lines are for $\lambda=1/20$, and blue dashed lines are for
$\lambda=20$. The corresponding lower one is for process
$|$a-d-b$\rangle$, while the upper one is for $|$b-f-c$\rangle$.
}\label{ent1D2D}
\end{figure}

We next consider two extreme cases of $\lambda \ll 1$ or $\lambda
\gg 1$ corresponding to the quasi-one- and quasi-two-dimensional
limits, respectively \cite{TCalarco}. The pair entanglement
results are shown in Fig. \ref{ent1D2D} for both $\lambda=1/20$
and $\lambda=20$. We find that the pair entanglement saturates to
a lower value in the quasi-one-dimensional limit at
$\lambda=1/20$, again because of the reduced motional state
degeneracy. Our results here are of course limited to the validity
regime of the model we adopt in the Hamiltonian (\ref{hm}). For a
more rigorous treatment, please refer to the methods developed in
Ref. \cite{jul} for harmonic traps or atomic waveguides.

\section{A toy model}
Much of our results above can in fact easily be appreciated from a
toy model for two distinguishable atoms, as described by the
Hamiltonian $H = H_0 + V$, with $H_0$ and $V$ given by
\begin{eqnarray}
H_0 &=& \hbar\omega (|10\rangle_{12}\langle 10|+ |01\rangle_{12}\langle 01|),\nonumber\\
V &=&  \hbar\delta |00\rangle_{12}\langle 00| \nonumber\\
&&+\hbar\eta(|10\rangle_{12}\langle 00|+|01\rangle_{12}\langle 00| + \rm{h.c.} ).
\label{toyh}
\end{eqnarray}
For simplicity, we have truncated the motional states to only
include the ground and the first excited state, denoted
respectively by $|0\rangle$ and $|1\rangle$ and
separated by the trap frequency $\omega$.
The collisional interaction is mainly limited to the ground
state manifold, giving rise to a level shift $\delta$
when both atoms are in the ground state ($|00\rangle$)
and a simple atom excitation with strength $\eta$.
Double excitation is assumed small and neglected.

\begin{figure}[htb]
\centering
\includegraphics[width=3.25in]{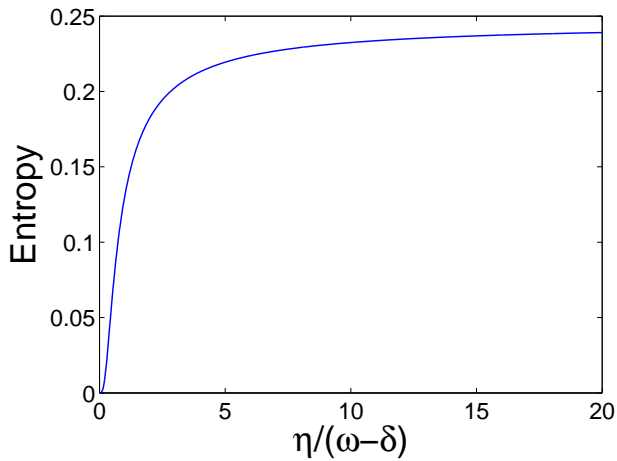}
\caption{(Color online) The dependence of pair entanglement as a
function of interaction parameter $\eta/(\omega-\delta)$.
}\label{toy}
\end{figure}

This toy model can be easily solved, and the pair entanglement as
function of interaction parameter $\eta/(\omega-\delta)$ is shown
in Fig. \ref{toy}. We see that the ground state is a simple
product state of each atom in the motional state $|0\rangle$ for
$\eta=0$, i.e., displaying no entanglement. With the increase of
atom-atom interaction, we find that the entanglement increases and
saturates to $2\ln (2)-{\sqrt{3}\over 2}\ln(2+\sqrt{3})\simeq
0.2458$, where the approximations for the toy become questionable.

It is important to note that this toy model reproduces the same
dependence of pair entanglement on the interaction strength as for
the system of two fermionic atoms during the BCS to BEC crossover.
We thus feel it is important to point out that from the point of
view of two atom motional state entanglement, nothing particularly
significant occurs during the BCS to BEC crossover.

\section{conclusion}
Before concluding, let us briefly summarize the generalization of
our current study to a narrow Feshbach resonance, where the bound
state molecular component in Eq. (\ref{2w}) cannot be neglected
anymore. The pair entanglement inside the molecular component can
be included by performing an analogous symmetric Schmidt
decomposition on Eq. (\ref{2w}). The total pair correlation can
then be computed analogously in terms of the entropy of the
independent Schmidt orbital expansion. Of course, such an approach
would require more details about the model formulation and the
exact determination of the molecular bound state wave function.

In summary, we have studied pair entanglement between two
spin-$1/2$ fermionic atoms inside a single optical lattice site
approximated by a cylindrical harmonic trap. We investigated
thoroughly the dependence of pair entanglement on the trap
strength and geometry and on atom-atom interaction strength along
the complete BCS to BEC crossover and focused on the broad
resonance regime. We developed a formalism for studying pair
entanglement including the effect of an effective range $r_0$ for
two interacting atoms at low energy. In the limit of a broad
Feshbach resonance, where the effect of $r_0$ becomes negligibly
small, our result reduces to the theory developed before for
evaluating pair entanglement in a single open channel of two atoms
without the presence of a bound molecule state. We find that pair
entanglement changes significantly against atomic pair interaction
because of the induced motional orbital deformations. In general,
however, the exact value of the spatial pair entanglement is also
governed by the motional state degeneracies. As a rule of thumb,
we find that spherical harmonic traps generally give rise to
larger pair entanglement. We hope our study will provide new
insights into the applications of quantum degenerate lattice
systems to quantum information science.

\section{acknowledgement}
We thank Dr. Peng Zhang for enlightening discussions. This work is
supported by CNSF and NSF.

%%%%%%%%%%%%%%%%%%%%%%%%%%%%%%%%%%%%%%%%%%%%%%%%%%%%%%%%%%%%%%%%%%%%%%%%%%%

\end{document}